\documentclass[preprint,3p, 11pt]{elsarticle}

\usepackage{amssymb}
\usepackage{amsmath}

\usepackage{amsmath,amssymb}
\usepackage{changepage}
\usepackage{textcomp,marvosym}
\usepackage{nameref,hyperref}
\usepackage[right]{lineno}
\usepackage[table]{xcolor}
\usepackage{array}
\usepackage{multirow}
\newcolumntype{+}{!{\vrule width 2pt}}
\newlength\savedwidth

\usepackage{caption}
\usepackage{booktabs}
\usepackage{lastpage,fancyhdr,graphicx}
\usepackage{epstopdf}
\usepackage{ragged2e}
\usepackage{tabularx}
\newcolumntype{C}{>{\centering\arraybackslash}X} 
\newcolumntype{L}{>{\raggedright\arraybackslash}X}
\justifying
\usepackage{setspace}
\usepackage{arydshln}
\biboptions{numbers,sort&compress}

\journal{}

\begin{document}

\begin{frontmatter}



\title{Multi-modal Data Fusion and Deep Ensemble Learning for Accurate Crop Yield Prediction}


\author[inst1]{Akshay Dagadu Yewle}

\affiliation[inst1]{organization={School of Computer Science and Informatics, Cardiff University}, addressline={Abacws, Senghennydd Road}, city={Cardiff}, postcode={CF24 4AG}, country={UK}.}

\author[inst1]{Laman Mirzayeva}

\author[inst1]{Oktay Karaku\c{s}}

\begin{abstract}
This study introduces \textit{RicEns-Net}, a novel Deep Ensemble model designed to predict crop yields by integrating diverse data sources through multimodal data fusion techniques. The research focuses specifically on the use of synthetic aperture radar (SAR), optical remote sensing data from Sentinel 1, 2, and 3 satellites, and meteorological measurements such as surface temperature and rainfall. The initial field data for the study were acquired through Ernst \& Young's (EY) Open Science Challenge 2023. The primary objective is to enhance the precision of crop yield prediction by developing a machine-learning framework capable of handling complex environmental data. A comprehensive data engineering process was employed to select the most informative features from over 100 potential predictors, reducing the set to 15 features from 5 distinct modalities. This step mitigates the ``curse of dimensionality" and enhances model performance. The RicEns-Net architecture combines multiple machine learning algorithms in a deep ensemble framework, integrating the strengths of each technique to improve predictive accuracy. Experimental results demonstrate that RicEns-Net achieves a mean absolute error (MAE) of 341 kg/Ha (roughly corresponds to 5-6\% of the lowest average yield in the region), significantly exceeding the performance of previous state-of-the-art models, including those developed during the EY challenge.
\end{abstract}




\end{frontmatter}



\section{Introduction}
This paper is grounded in the purpose and drive behind one of the Sustainable Development Goals (SDG) outlined by the United Nations—a comprehensive set of 17 objectives to be achieved by 2030 \cite{UNSDG}. These goals collectively embody humanity’s pursuit of a sustainable future for both the planet and its inhabitants. Serving as a global framework, the 17 SDGs guide international endeavours to address the challenges posed by climate change while harmonising human ambitions for prosperity and improved quality of life. Central to this vision is the assurance of food security \cite{UNSDGGoal2}, particularly for a significant portion of the global population residing in environmentally vulnerable areas susceptible to the impacts of climate and weather fluctuations.

Rice is one of the most important staple foods globally, feeding more than half of the world’s population. It is cultivated on over 160 million hectares, producing around 500 million metric tons annually, with the majority of production concentrated in Asia. The region accounts for nearly 90\% of global rice output, with countries such as China, India, and Indonesia leading in production. Besides being a dietary staple, rice also supports the livelihoods of millions of farmers, playing a vital role in rural economies. However, rice cultivation is highly resource-intensive, requiring significant amounts of water and labour, and is particularly vulnerable to climate change. Rising temperatures, changing precipitation patterns, and the increasing frequency of extreme weather events threaten rice yields globally, posing challenges to food security in many regions.

This study underscores the critical importance of rice crops and Vietnam, focusing on their global and regional significance. Rice, often referred to as a “Gift of God” for its nutritional value, is a cornerstone of food security and public health worldwide. Despite its critical role, less than 8\% of global rice production enters international trade \cite{FAO:2022}, emphasizing its largely localized consumption patterns, as noted by the Food and Agriculture Organisation Corporate Statistical Database (FAOSTAT). However, Vietnam stands as an exception, being one of the world’s leading rice exporters. The Mekong Delta, in particular, is central to the nation’s status as a global rice powerhouse, with its fertile paddy fields supporting diverse rice varieties.

In Vietnam, rice holds unparalleled cultural, economic, and ecological importance. It serves as a dietary staple, forming the foundation of traditional cuisine, while also being integral to the livelihoods of millions of rural farmers. The sector is a driver of economic development, supported by government initiatives aimed at boosting productivity and sustainability. However, the ecological and climatic demands of rice cultivation present significant challenges. As a water-intensive crop, rice is highly sensitive to temperature fluctuations, requiring an optimal range of 25–35°C. Brief exposure to extreme heat or untimely rainfall during the reproductive stage can result in sterility or yield loss. This vulnerability, coupled with the long growing periods, makes rice cultivation particularly susceptible to climate change, underscoring the urgency of sustainable practices in Vietnam and globally.

This paper introduces \textit{RicEns-Net}, a novel ensemble learning model designed for accurate crop yield prediction. By integrating five diverse remote sensing data sources—Sentinel-1, Sentinel-2, Sentinel-3, NASA’s Goddard Earth Sciences (GES) Data and Information Services Center (DISC), and field measurements—the proposed framework leverages sophisticated data engineering techniques to enhance predictive performance. The model effectively combines synthetic aperture radar (SAR), multispectral imaging (MSI), meteorological parameters, and ground-truth observations, ensuring a robust and comprehensive approach to crop yield estimation. The inclusion of multi-modal data sources helps mitigate uncertainties associated with individual datasets, improving both spatial and temporal resolution in predictions.

The remainder of this paper is organized as follows: Section \ref{sec:related} reviews related work, analysing previous findings and establishing the foundation for the study’s objectives and contributions. Section \ref{sec:methods} details the materials and methods, including descriptions of data sources, preprocessing techniques, and feature engineering approaches. Section \ref{sec:results} presents the experimental results derived from the proposed model, followed by Section \ref{sec:disscussions}, which interprets these findings and discusses their implications. Finally, Section \ref{sec:conc} summarizes the study, outlines its limitations, and highlights potential directions for future research.

\section{Related Works}\label{sec:related}
Advancements have influenced the evolution of crop yield monitoring in sensor technology and data analysis methodologies. While early studies highlighted rainfall as the primary determinant of crop yield, a key shift occurred in 1968 with the recognition of soil moisture as a more reliable predictor by Baier and Robertson \cite{baier1968performance}. Their work leverages spectral data to estimate crop yield based on vegetation health indicators. Over time, numerous vegetation indices (VIs) have been developed to assess vegetative conditions and physiological characteristics of crops. These VIs, including the Normalized Difference Vegetation Index (NDVI), Leaf Area Index (LAI) \cite{bouman1995crop}, and Transformed Soil Adjusted Vegetation Index (TSAVI) \cite{baret1991potentials}, play a crucial role in crop prediction models. Advancements in hyperspectral imaging have enabled the capture of fine-grained spectral data, facilitating the development of biochemical indices for quantifying plant constituents.

Rice crop yield estimation relies on understanding the crop's growth stages and environmental factors. Water level in paddy fields, rather than direct precipitation, is crucial for irrigated rice fields. Accumulating temperature is more important than temperature at certain times, as it affects the crop's development stages. These factors are integrated into crop models to predict yield accurately.

In the early 2000s, research surged leveraging imaging and machine learning technologies for crop yield prediction. Studies introduced novel methodologies, such as artificial neural networks (ANN) and SVR, to analyse remote sensing data and historical yield records. These approaches demonstrated more precise results compared to traditional models and prepared for more precise and scalable methods for estimating crop yields. Uno et al. \cite{uno2005artificial} analyse hyperspectral images of corn plots in Canada using statistical and ANN approaches, demonstrating the potential of ANNs in predicting yield with higher accuracy compared to conventional models. Li et al. \cite{li2007estimating} introduce a methodology employing ANN models to predict corn and soybean yields in the United States "corn belt" region, achieving high prediction accuracy through historical yield data and NDVI time series. Bala and Islam \cite{bala2009correlation} estimate potato yields in Bangladesh using TERRA MODIS reflectance data and Vegetation Indices (VIs), demonstrating the effectiveness of VIs derived from remote sensing for early yield estimation. Li et al. \cite{li2009winter} employ SVR and multi-temporal Landsat TM NDVIs to predict winter wheat yield in China, showcasing the precision and effectiveness of SVR models in yield estimation. Stojanova et al. \cite{stojanova2010estimating} integrate LiDAR and Landsat satellite data using machine learning techniques to model vegetation characteristics in Slovenia. Their approach combines the precision of LiDAR data with the broad coverage of satellite data, facilitating effective forest management and monitoring processes.

Furthermore, Mosleh et al. \cite{mosleh2015application} evaluated the efficacy of remote sensing imagery in mapping rice areas and forecasting production, highlighting challenges such as spatial resolution limitations and issues with radar imagery. Johnson et al. \cite{johnson2016crop} developed crop yield forecasting models for the Canadian Prairies, revealing the effectiveness of satellite-derived vegetation indices, particularly NDVI, in predicting yield potential. Pantazi et al. \cite{pantazi2016wheat} proposed a model for winter wheat yield prediction, integrating soil spectroscopy and remote sensing data to visually depict yield-influencing factors. Ramos et al. \cite{ramos2020random} introduced an optimised Random Forest algorithm for maize-crop yield prediction, emphasising the importance of vegetation indices like NDVI, NDRE, and GNDVI. Li et al. \cite{li2021machine} utilised extreme gradient boosting machine learning to accurately predict vegetation growth in China, achieving high predictive accuracy and demonstrating effectiveness under diverse conditions. Zhang et al. \cite{zhang2021integrating} employed field-surveyed data to predict smallholder maize yield, with novel insights into the performance of various vegetation indices and machine learning techniques.

Recent studies have demonstrated the efficacy of utilising Sentinel-2 satellite imagery and machine learning techniques for predicting crop yields and mapping crop types in various agricultural settings. Son et al. \cite{son2022field} employed Sentinel-2 image composites and machine learning algorithms to forecast rice crop yields in Taiwan, finding that Support Vector Machines (SVM) outperformed RF and ANN at the field level, indicating their potential for accurate yield predictions approximately one month before harvesting. Perich et al. \cite{perich2023pixel} utilised Sentinel-2 imagery to map crop yields at the pixel level in small-scale agriculture, with machine learning models utilising spectral index and raw reflectance data proving effective, even in the presence of cloudy satellite time series. Khan et al. \cite{khan2023early} combined ground-based surveys with Sentinel-2 satellite images and deep learning techniques to map crop types, achieving high accuracy in identifying staple crops like rice, wheat, and sugarcane within the first four weeks of sowing. 

Along with Sentinel 2, UAV and other sensor spectral information have also been used in the literature in the last couple of years. Shafi et al. \cite{shafi2023tackling} propose XGBoost, LASSO and RF regression models to be utilised via Drone-based multispectral imagery whilst Islam et al. \cite{islam2023rapid} combine remote sensing and meteorological data in stacking multiple regression models for rice crop yield prediction. Zhou et al. \cite{zhou2023rice} compare CNN and LSTM-based models for predicting annual rice yield in Hubei Province, China, utilising ERA5 temperature data, and MODIS vegetation indices, demonstrating that the CNN-LSTM model with spatial heterogeneity outperforms models using only remote sensing data. Arshad et al. \cite{arshad2023applicability} evaluate the performance of RF and SVR, in predicting wheat yield in southern Pakistan using a combination of remote sensing indices and climatic variables where RF outperforms other methods. Asadollah et al. \cite{asadollah2024optimizing} assess the effectiveness of using a novel Randomized Search cross-validation (RScv) optimization algorithm with four machine learning models to predict annual yields of four crops (Barley, Oats, Rye, and Wheat) across 20 European countries, demonstrating improved prediction accuracy through satellite-based climate and soil data. 

Furthermore, Lu et al. \cite{lu2024goa} present a state-of-the-art CNN-BiGRU model enhanced by GOA and a novel attention mechanism (GCBA) for accurate county-level soybean yield estimation in the U.S., leveraging multi-source remote sensing data and outperforming existing models in yield prediction accuracy. Killeen et al. \cite{killeen2024corn} investigate UAV-based corn yield prediction using RF and linear regression models and find that spatial cross-validation reduces over-optimism in yield prediction compared to standard 10-fold cross-validation, with LR showing better spatial generalizability than RF. Dhaliwal and Williams II \cite{dhaliwal2024sweet} use a 26-year dataset on US sweet corn production to evaluate machine learning models for yield prediction, finding that RF performs best, with year, location, and seed source identified as the most influential variables.

Recently, Gadupudi et al. \cite{gadupudi2024adaptive} demonstrated integrating ML strategies like RF and Decision Trees alongside DL models such as LSTM and RNN to enhance crop prediction accuracy, incorporating soil attributes, climate data, and cost analyses to optimize outcomes. Similarly, Rao et al. \cite{rao2024brinjal} employed attention-based CNNs and bidirectional LSTMs with hyperparameter tuning to predict crop yields, showcasing significant improvements in detection performance through methods like the shuffling shepherd optimization algorithm. Sharma et al. \cite{sharma2025enhancing} explored the fusion of AI algorithms, including logistic regression and IoT-enabled analytics, to tailor recommendations based on regional agricultural parameters, advancing productivity and diversification. 

The diverse literature outcomes examined previously, along with numerous others, have highlighted the multidimensional potential of AI in transforming traditional agricultural practices into more data-driven, adaptive systems. They have employed a variety of data types from different sources and machine learning models. This diversity presents challenges in generalising techniques across different datasets, yet it also enhances performance for specific datasets. The adoption of multi-modal data usage, multi-modal AI techniques, and Ensemble methods has emerged as the current practice in this research field. Shahhosseini et al. \cite{shahhosseini2021corn} explore the predictive performance of two novel CNN-DNN machine learning ensemble models for forecasting county-level corn yields across the US Corn Belt. By combining management, environmental, and historical yield data from 1980 to 2019, the study compares the effectiveness of homogeneous and heterogeneous ensemble creation methods, finding that homogeneous ensembles provide the most accurate yield predictions, offering the potential for the development of a reliable tool to aid agronomic decision-making. Gavahi et al. \cite{gavahi2021deepyield} introduce DeepYield, a novel approach for crop yield forecasting that combines Convolutional Long Short-Term Memory (ConvLSTM) and 3-Dimensional CNN (3DCNN). By integrating spatiotemporal features extracted from remote sensing data, including MODIS Land Surface Temperature (LST), Surface Reflectance (SR), and Land Cover (LC), DeepYield outperforms traditional methods and demonstrates more precise forecasting accuracy for soybean yields across the Contiguous United States (CONUS). Zare et al. \cite{zare2024within} investigate the impact of data assimilation techniques on improving crop yield predictions by assimilating LAI data into three single crop models and their multimode ensemble using a particle filtering algorithm. Results from a case study in southwestern Germany reveal that data assimilation significantly enhances LAI simulation accuracy and grain yield prediction, particularly for certain crop models, highlighting the potential for further improvements in data assimilation applications through regional model calibration and input uncertainty analysis.

Moreover, Gopi and Karthikeyan \cite{gopi2024red} introduce the Red Fox Optimization with Ensemble Recurrent Neural Network for Crop Recommendation and Yield Prediction (RFOERNN-CRYP) model, which leverages deep learning methods to provide automated crop recommendations and yield predictions. By employing ensemble learning with three different deep learning models (LSTM, bidirectional LSTM (BiLSTM), and gated recurrent unit (GRU)) and optimising hyperparameters using the RFO algorithm, the proposed model demonstrates improved performance compared to individual classifiers, offering valuable support for farmers in decision-making processes related to crop cultivation. From a similar perspective, Boppudi et al. \cite{boppudi2024deep} propose a deep ensemble model for accurately predicting crop yields in India, addressing the challenge posed by variations in weather and environmental factors. The model (Deep Max Out, Bi-GRU and CNN) incorporates improved preprocessing techniques, feature selection using the IBS-BOA algorithm, and prediction through a combination of Deep Ensemble Model and Ensemble classifiers, resulting in significantly reduced error rates compared to existing methods.

Recently, Umamaheswari and Madhumathi \cite{umamaheswari2024predicting} applied a stacking ensemble approach using regressors like SVR, KNN, and RF as base learners, with LASSO regression as the meta-learner, achieving enhanced prediction precision. Osibo et al. \cite{osibo2024integrating} demonstrated the integration of weighted ensemble methods with remote sensing data, achieving better results compared to state-of-the-art models while simplifying data integration. Zhang et al. \cite{zhang2024ensemble} constructed the StackReg framework, combining UAV-acquired multispectral data with ridge regression, SVM, Cubist, and XGBoost, which consistently outperformed base models, particularly in multi-stage settings. These studies highlight the critical role of ensemble learning in improving prediction reliability, offering versatile solutions adaptable to diverse agricultural contexts.

In the sequel, we introduce a novel framework for predicting crop yields, named ``\textit{RicEns-Net}." This framework incorporates advanced data engineering processes involving five unique data sources, namely Sentinel 1/2/3, NASA's Goddard Earth Sciences (GES) Data and Information Services Centre (DISC), and field measurements. The novelty of RicEns-Net lies in its integration of these diverse and rich sources of multi-modal data, comprising 15 features selected from a pool of over 100, within an advanced Deep Ensemble model. This model encompasses widely-used CNN and MLP architectures, as well as less explored DenseNet and Autoencoder architectures, which have been infrequently utilised or entirely unexplored in existing literature.

\subsection{Objectives \& Contributions}
The key objectives and related contributions of this research are as follows:
\begin{itemize}
    \item Integrate diverse remote sensing and meteorological data to enhance the accuracy and reliability of crop yield forecasting.
    
    \textit{Contribution:} We successfully unified radar, optical imagery, and meteorological data into a coherent multimodal dataset. This integration demonstrated the practical benefits of combining diverse data sources, providing a foundation for robust and accurate crop yield forecasting models.
    
    \item Address feature complexity and dimensionality challenges through advanced feature engineering and selection.
    
    \textit{Contribution:} We developed and implemented novel feature engineering and selection techniques, effectively reducing feature dimensionality while preserving essential predictive attributes. This contribution advanced the methodology for handling multimodal datasets in agricultural forecasting.
    
    \item Develop a novel deep ensemble learning model to improve the precision of crop yield prediction using multimodal data sources.
    
    \textit{Contribution:} We designed a deep ensemble learning framework that leverages data from SAR, MSI, and meteorological sources. The model achieved state-of-the-art performance, significantly enhancing precision and reliability in crop yield prediction tasks.
    
    \item Demonstrate the effectiveness of multimodal data fusion through performance comparisons with state-of-the-art machine learning techniques.
    
    \textit{Contribution:} We performed extensive benchmarking of the proposed model against leading machine learning techniques, validating the effectiveness of multimodal data fusion. The results highlighted the higher performance of the model and provided insights into the role of data integration in improving predictive accuracy.
\end{itemize}

\section{Materials and Methods}\label{sec:methods}
\subsection{Study Area \& Rice Crop Details}

\begin{figure}[t!]
\centering
\includegraphics[width=\linewidth]{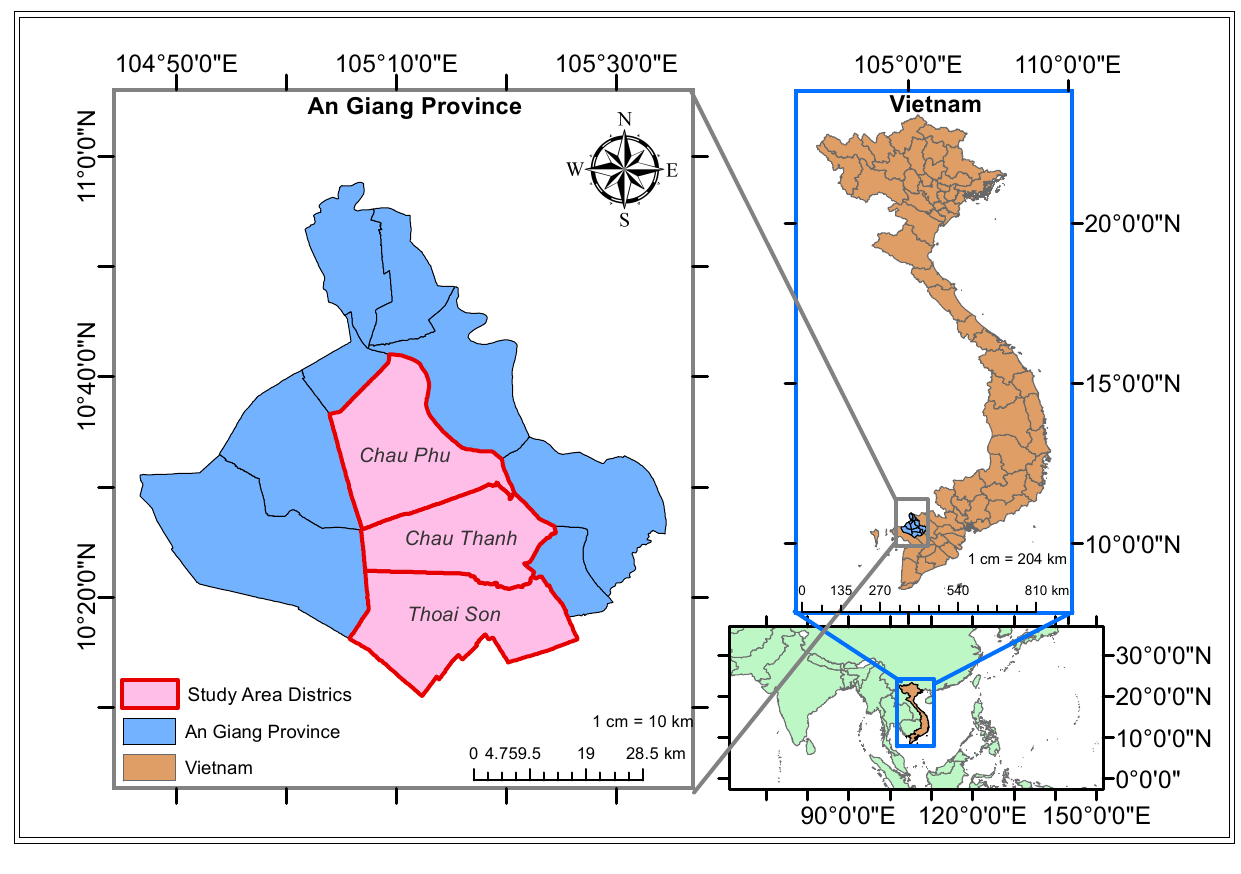}
\caption{Map illustrating the geographical area under investigation, encompassing the An Giang province in Vietnam, including the specific study districts of Chau Thanh, Thoai Son, and Chau Phu. This region is significant for its rice production and diverse environmental conditions, making it an ideal case study for testing the generalizability of the RicEns-Net model.}
\label{fig:vietnam}
\end{figure}

As stated earlier, this study begins by employing the dataset offered by EY for the 2023 iteration of their Open Science Data Challenge \cite{EY2023}. The dataset encompasses information from 557 farm sites situated in Chau Thanh, Thoai Son, and Chau Phu districts within the province of An Giang in Vietnam (see Figure \ref{fig:vietnam}). The study province of An Giang relies significantly on agriculture as a cornerstone of its economy. Notably, An Giang province is situated in the Mekong River delta region, crucial for providing irrigation to support rice cultivation. The dataset, supplied by EY, contains fundamental details for each crop, including District Name, Latitude, Longitude, Crop Season, Crop Frequency, Harvest Date, Crop Area, and Yield as given in Table \ref{tab:vietnam}.

\begin{table}[ht!]
\centering
\caption{Study Area Details, including district names, geographic locations, population, area, and population density for Chau Thanh, Thoai Son, and Chau Phu districts in An Giang province, Vietnam. This information provides context for the study region's significance in rice production.}
\begin{tabularx}{\textwidth}{C p{1.75cm} p{1cm} C C p{1.5cm} C}
\toprule
\textbf{District} & \textbf{Province} & \textbf{Data Count} & \textbf{Geographic Location} & \textbf{Population} & \textbf{Area} & \textbf{Population Density} \\
\toprule
Chau Thanh & & 218 & & 130,101 & 571 km$^2$ & 228 /km$^2$ \\
\cline{1-1} \cline{3-7}
Thoai Son & An Giang & 171 & Mekong Delta & 187,620 & 456 km$^2$ & 411 /km$^2$ \\
\cline{1-1} \cline{3-3} \cline{5-7}
Chau Phu & & 168 & Region & 250,567 & 426 km$^2$ & 588 /km$^2$ \\
\bottomrule
\end{tabularx}
\label{tab:vietnam}
\end{table}

Every entry in the primary dataset represents an individual crop and is characterised by eight features, including three categorical variables (District name, Crop Season [WS=Winter Spring; SA = Summer Autumn], and Crop Frequency of the specified farm [D = Twice; T = Thrice]) and five numerical variables (Latitude, Longitude, Harvest Date, Area [Hectares], and Yield Rate [Kg/Ha]). The harvesting dates of the crops in question extend from March 18, 2022, to August 9, 2022, covering two significant crop seasons, namely Summer-Autumn and Winter-Spring.

\begin{figure}[t!]
\centering
\includegraphics[width=\linewidth]{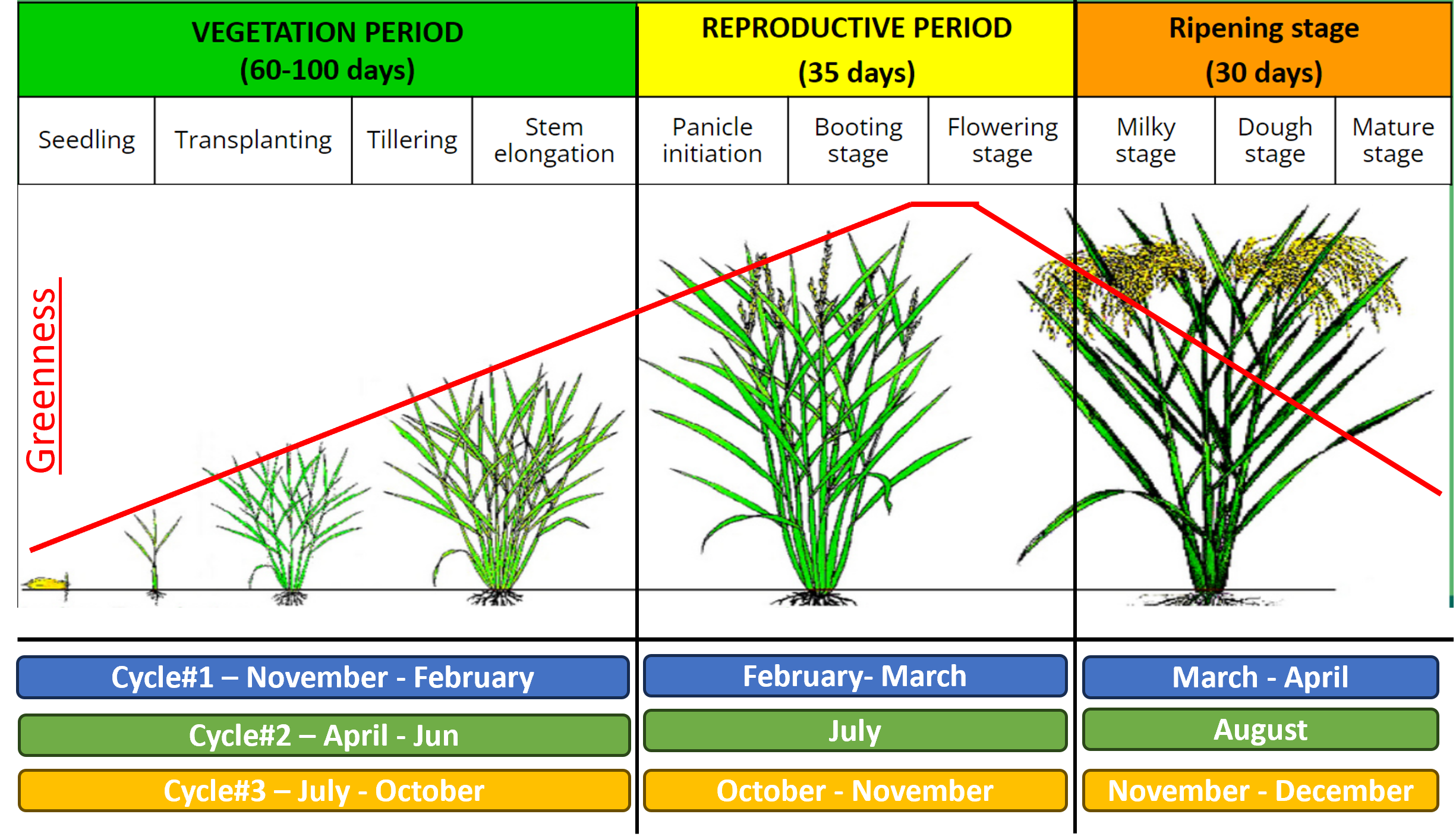}
\caption{Rice growing stages and three potential crop cycles in Vietnam's study region. Credit: \cite{irri-extension}}
\label{fig:regions}
\end{figure}

While the overall duration of the crop cycle spans approximately 5-6 months, contingent on the season, the period from transplanting to harvesting typically ranges from 90 to 130 days. The growth trajectory can be categorised into three principal phases: Vegetative, Reproductive, and Ripening stages, illustrated in Figure \ref{fig:regions}. During the Reproductive stage, rice plants attain their maximum greenness, marking the culmination of this phase. Subsequently, the crop transitions into the Ripening stage, characterised by the transformation of the plants from green to yellow, coinciding with the maturation of rice grains. In the context of Vietnam, rice cultivation occurs biannually or tri-annually within two seasons: Winter-Spring (Nov-Apr), Summer-Autumn (Apr-Aug), and Autumn-Winter (Jul-Dec).

The necessary geospatial information can be obtained from either Landsat or Sentinel satellites, taking into account the designated location and harvest time outlined in the dataset. After careful consideration of technical details such as Ground Spatial Resolution (GSD) and revisit frequency, the decision is made to opt for data sourced from Sentinel satellites. Additionally, there is anticipation in leveraging the complete spectrum of SAR, MSI, and Meteorological data. To gather remote sensing data for a specified area, it is essential to finalise two key parameters: the time window and the geographical bounding box or crop window.

We intend to gather data during the phase when the crop is nearing complete maturation. The progress of the crop's growth is discerned through the intensity of the green hue in the data, with the period from transplantation to maturity spanning 60-100 days, contingent on the prevailing season. Following full maturity, the crop progresses into the ripening stage, which typically lasts around 30 days. To determine the timeframe for data collection, we use the harvest date as a reference point. The initiation date is established as 60-90 days before the harvest date, while the conclusion date is set at 30 days before the harvest date (refer to Figure \ref{fig:regions}). This designated time window encapsulates the entire duration from the crop's transplantation to the conclusion of the maturity stage, ensuring that the remote sensing data aligns with the various growth phases of the crop.

The primary input data exclusively provides information about the field's area without specifying the precise bounding region for each field. This lack of detail hinders our ability to extract data for the exact boundaries of individual crop fields. To address this limitation, we proceed by identifying the precise location of each field in the MSI data and cropping the image to obtain a set of pixels surrounding that specific location. Given that the visual bands of the MSI data have a spatial resolution of 10 meters per pixel, a 3x3 pixel collection corresponds to an area of 900 square meters, equivalent to 0.222 acres or 0.09 hectares. Notably, the SWIR16 and SWIR22 bands of Sentinel 2's MSI exhibit a spatial resolution of 20 meters per pixel, meaning a 3x3 pixel bounding box represents an area of 3600 square meters, or 0.89 acres, or 0.36 hectares.

\subsection{Data/Feature Extraction}
\subsubsection{Sentinel-1}
Sentinel-1 operates with four distinct acquisition modes: Stripmap (SM), Interferometric Wide Swath (IW), Extra Wide Swath (EW), and Wave (WV). During Synthetic Aperture Radar (SAR) acquisition, known as the “datatake” process, data from SM, IW, and EW modes is divided into smaller, manageable slices. These slices undergo processing to generate different product levels. Level-0 products contain raw sensor data, while Level-1 products provide calibrated imaging data such as Single Look Complex (SLC) and Ground Range Detected (GRD) formats, essential for interferometry and georeferenced analysis. Finally, Level-2 products include higher-level geophysical measurements, such as Ocean (OCN) products, which support oceanographic and environmental applications.

Sentinel 1 delivers SAR images featuring two polarisations, vertical (VV) and horizontal (VH), characterised by the difference in the polarisation of their transmitted and received signals. The polarisation of radar signals plays a crucial role in deciphering the structure and orientation of surface elements on the land. The radar signal experiences scattering and depolarisation due to the randomly oriented structure of plant leaves as it undergoes multiple bounces. By comparing the vertical (VV) and horizontal (VH) components, the degree of scattering by the land surface can be discerned. In our model, we incorporate this technique through the $VV/VH_{ratio}$ feature. The data was collected by defining a geographical bounding box with a size of 0.001 degree, resulting in an output array size of 3x3. However, some location data did not conform to this shape and required trimming to achieve a 3x3 box size. Given Sentinel 1's spatial resolution of 10 meters, each box corresponds to an area of 30 meters by 30 meters. The subsequent features were derived from Sentinel 1 SAR data 
\begin{itemize}
\item Set 1 (4 variables) $\rightarrow$ $VV_{mean}, VH_{mean}, VV/VH_{ratio,mean}, RVI_{mean}$
\end{itemize}
where the radar vegetation index (RVI) is given as
\begin{align}
RVI = \dfrac{VV}{VV + VH}.
\end{align}

\subsubsection{Sentinel-2}
Sentinel-2 imagery is collected through a continuous acquisition process known as ‘datatake,’ covering up to 15,000 km in length. The acquired data is structured into products at different processing levels. Sentinel-2 products are organized into tiles or granules, following a Universal Transverse Mercator (UTM) projection system to ensure global coverage.

At Level-1B, the data is provided in granules of 23 km × 25 km, containing radiometrically calibrated and geometrically refined imagery. The Level-1C and Level-2A products are structured into 110 km × 110 km tiles, ensuring seamless coverage across designated UTM zones. Each tile overlaps with neighboring ones to maintain spatial continuity and facilitate multi-temporal analysis. Level-1C includes top-of-atmosphere (TOA) reflectance, while Level-2A applies atmospheric corrections to generate bottom-of-atmosphere (BOA) reflectance, making it suitable for land surface monitoring and vegetation analysis.

Sentinel 2 data furnishes spectral intensities across 13 bands, encompassing Visual-NIR (VNIR) to Short-Wave Infra-Red (SWIR) regions. Notably, there are four spectral bands, namely Red, Green, Blue, and NIR (B04, B03, B02, and B08), offering a ground resolution of 10 meters. Additionally, six bands exhibit a 20-meter ground resolution, comprising four Red Edge bands (B05, B06, B07, and B08A) and two SWIR bands with distinct wavelengths (B11 and B12). The remaining three bands, with a 60-meter ground resolution, serve specific purposes: B01 for aerosol detection (0.443 $\mu$m), B09 for water vapour observation (0.945 $\mu$m), and B10 for cirrus detection (1.374 $\mu$m). Notably, the Sentinel 2 mission boasts a revisit frequency of 5 days.

When acquiring MSI data, it is crucial to account for the potential impact of cloud cover in the targeted area. With a revisit frequency of 5 days, there are only 6-8 chances to capture images during the period when crops reach full growth before maturation. Given Vietnam's tropical monsoon climate, these image opportunities are prone to cloudiness. The dataset at hand reveals median cloud coverage values of 16\% and 21\% during the Winter-Spring and Summer-Autumn seasons, respectively.

In order to prevent the occurrence of unclear or cloudy images, it is necessary to eliminate those with a high level of cloud coverage. Simultaneously, we aim to capture comprehensive crop data when the plants are at their full growth and exhibit maximum greenness. To achieve optimal outcomes, we conducted experiments with various values for maximum cloud coverage and time windows. Based on the findings detailed in Table \ref{tab:cloud}, we concluded that setting the maximum cloud coverage to 60\% and collecting Sentinel-2 MSI data during the 50 days preceding the crop maturation provides favourable results. The objective is to secure a minimum of 4-5 images for each specific location.

\begin{table}[ht!]
\renewcommand{\arraystretch}{0.5}
\centering
\caption{Trials For Identifying Optimal Value Of Cloud Coverage Threshold \& Time Window For Data Collection}
\begin{tabularx}{\textwidth}{p{0.75cm}p{1.5cm}p{1.5cm}CCCCCCCp{1.5cm}}
\toprule
\textbf{Trial} & \textbf{Max.} & \textbf{Window} & \multicolumn{7}{c}{\textbf{No. Of Crops}} & \textbf{Remarks}\\
\cmidrule{4-10} & \textbf{Coverage} & \textbf{(\textit{pre-}Maturity)} & 0&1&2&3&4&5&$>$5 & \\
\midrule
\textbf{1} & 25\% & 30 & 119 & 298 & 138 & 0 & 2 & 0 & 0 & \textcolor[rgb]{ 1, 0, 0}{\textbf{Reject}} \\
\midrule
\textbf{2} & 30\% & 30 & 119 & 298 & 138 & 0 & 0 & 2 & 0 & \textcolor[rgb]{ 1, 0, 0}{\textbf{Reject}} \\
\midrule
\textbf{3} & 40\% & 30 & 13 & 348 & 83 & 109 & 2 & 0 & 2 & \textcolor[rgb]{ 1, 0, 0}{\textbf{Reject}} \\
\midrule
\textbf{4} & 25\% & 40 & 5 & 412 & 138 & 0 & 2 & 0 & 0 & \textcolor[rgb]{ 1, 0, 0}{\textbf{Reject}} \\
\midrule
\textbf{5} & 25\% & 45 & 5 & 380 & 170 & 0 & 2 & 0 & 0 & \textcolor[rgb]{ 1, 0, 0}{\textbf{Reject}} \\
\midrule
\textbf{6} & 25\% & 50 & 5 & 358 & 114 & 78 & 0 & 2 & 0 & \textcolor[rgb]{ 1, 0, 0}{\textbf{Reject}} \\
\midrule
\textbf{7} & 30\% & 50 & 5 & 252 & 204 & 94 & 0 & 0 & 2 & \textcolor[rgb]{ 1, 0, 0}{\textbf{Reject}} \\
\midrule
\textbf{8} & 40\% & 50 & 0 & 250 & 204 & 94 & 0 & 2 & 2 & \textcolor[rgb]{ 1, 0, 0}{\textbf{Reject}} \\
\midrule
\textbf{9} & 50\% & 50 & 0 & 0 & 57 & 193 & 0 & 15 & 287 & \textcolor[rgb]{ 1, 0, 0}{\textbf{Reject}} \\
\midrule
\textbf{10} & \textcolor[rgb]{ 0, .69, .314}{\textbf{60\%}} & 50 & 0 & 0 & 0 & 0 & 70 & 200 & 287 & \textcolor[rgb]{ 0, .69, .314}{\textbf{Accept}} \\
\bottomrule
\end{tabularx}
\label{tab:cloud}
\end{table}

To ensure that the spectral intensity trends are captured, we identify the minimum, maximum, mean and variance of 9 MSI bands based on all the MSI images available for each location.
\begin{itemize}
\item Set 2 (36 variables) $\rightarrow$ is in a format of ``$Band_{stats}$" where $Band = \{B02$, $B03$, $B04$, $B05$, $B06$, $B07$, $B08$, $B11$, $B12\}$ and $stats = \{min$, $max$, $mean$, $var\}$.
\end{itemize}

MSI data has been used to create transformational features known as \textit{Vegetation Indices} given in Table \ref{tab:indices} like NDVI, SR, EVI, EVI2, SAVI, RGVI, DVI, MSR, NIRv, kNDVI, NDVIre, NDRE1, NDRE2 to indicate the volume of vegetation on the land surface.
\begin{itemize}
\item Set 3 (26 variables) $\rightarrow$ is in a format of ``$VI_{stats2}$" where $VI = \{NDVI$, $SR$, $EVI$, $EVI2$, $SAVI$, $RGVI$, $DVI$, $MSR$, $NIRv$, $kNDVI$, $NDVIre$, $NDRE1$, $NDRE2\}$ and $stats2 = \{mean$, $var\}$.
\end{itemize}

Utilising MSI data, various features, such as NDWI, BSI, and LSWI, as outlined in Table \ref{tab:indices}, have been generated to depict soil and water content. This application is particularly advantageous in the context of rice cultivation, where the crop is submerged in water.
\begin{itemize}
\item Set 4 (6 variables) $\rightarrow$ is in a format of ``$VI2_{stats2}$" where $VI2 = \{NDWI$, $BSI$, $LSWI\}$ and $stats2 = \{mean$, $var\}$.
\end{itemize}

Additional features have been generated using MSI spectral data, incorporating information derived from the biochemical properties of the plants (refer to Table \ref{tab:indices}).
\begin{itemize}
\item Set 5 (4 variables) $\rightarrow$ is in a format of ``$VI3_{stats2}$" where $VI3 = \{CCI$, $GCC\}$ and $stats2 = \{mean$, $var\}$.
\end{itemize}

\subsubsection{Sentinel-3}
Sentinel-3 ensures the continuous availability of high-quality data for monitoring land, ocean, and atmospheric conditions, particularly in coastal areas where accuracy is critical. The mission provides comprehensive environmental observations globally, supporting a range of applications. Sentinel-3 plays a key role in fire detection, inland water surface height measurements, and land ice/snow surface temperature assessments. Its multi-instrument payload enables precise monitoring of ocean colour, sea surface temperature, and land surface dynamics, contributing to climate research, water resource management, and disaster response.

Sentinel 3 data was acquired to obtain meteorological information related to environmental variables such as ambient air temperature (MET$_{\text{temp}}$), land surface temperature (LST), solar radiation (MET$_{\text{solrad}}$), and specific humidity (MET$_{\text{sh}}$). These data sets have been integrated into the model as the following features:
\begin{itemize}
\item Set 6 (8 variables) $\rightarrow$ is in a format of ``$S3_{stats2}$" where $S3 = \{MET_{temp}$, $LST$, $MET_{solrad}$, $MET_{sh}\}$ and $stats2 = \{mean$, $var\}$.
\end{itemize}

\subsubsection{NASA GES DISC}
Rainfall information was acquired from NASA's Goddard Earth Sciences (GES) Data and Information Services Centre (DISC) through the utilisation of the Google Earth Engine API. The data retrieval involved the utilisation of the \textit{precipitationCal} parameter, which denotes rainfall in mm per hour. We organise this data into two distinct features: Rainfall-Total$_{\text{growth}}$ and Rainfall-Total$_{\text{maturity}}$, encompassing three statistical measures—mean, maximum, and sum as
\begin{itemize}
\item Set 7 (6 variables) $\rightarrow$ is in a format of ``$NASAGES_{stats3}$" where $NASAGES = \{Rainfall_{growth}$, $Rainfall_{maturity}\}$ and $stats3 = \{mean$, $max$, $sum\}$.
\end{itemize}

\subsection{Correlation Analysis}
After meticulously gathering all potentially valuable engineered features from multi-modal remote sensing data, we proceed to examine their statistical and predictive capabilities for subsequent feature selection. This phase, delineated in this sub-section, initiates with a correlation analysis.

Concerning the relationship between SAR features and $Yield_{rate}$, all four data features exhibit a strong correlation with $Yield_{rate}$. As anticipated, the $VH_{mean}$ feature effectively captures the backscattering of the SAR signal by the rice plant leaves, resulting in a higher correlation (0.32) with $Yield_{rate}$ compared to $VV_{mean}$ (0.25). The $VV/VH_{ratio}$ serves as a transformative feature, demonstrating an enhanced correlation (0.45) in comparison to both $VV_{mean}$ and $VH_{mean}$ individually. Notably, the Radar Vegetation Index (RVI) shows a similar positive correlation (0.41) with the Yield Rate.

As previously indicated regarding Sentinel-2 data, we derive spectral statistics from 9 bands: B02 Blue, B03 Green, B04 Red, B05-B07 Red Edge, B08 NIR, and B11-B12 SWIR. These statistics, namely min, max, mean, and variance, are incorporated into the model. Upon analysing observations across all bands, it is noteworthy that variance features exhibit a relatively low correlation ($< 0.1$) with the target variable, whereas other statistical measures demonstrate correlation coefficient values surpassing 0.3.

Spectral data is employed to generate transformative characteristics known as Vegetative Indices (VI). These indices serve as a more efficient measure for discerning and monitoring variations in plant phenology. In our approach, we utilise VIs such as NDVI, NDVIre, NDRE1, NDRE2, SR, DVI, MSR, EVI, EVI2, SAVI and RGVI, NIRv, and kNDVI. These features are integrated into the model in the form of their respective mean and variance features. However, the variance feature is omitted from the model due to its limited correlation with the target variable. Notably, kNDVI exhibits one of the highest correlations with Yield, while features like DVI, EVI, and NIRv demonstrate some of the lowest correlations.

Similar to the vegetation indices, we can employ optical data to compute additional indices that precisely quantify the environmental conditions of the crop's cultivation. NDWI, LSWI16, and LSWI20 specifically indicate the water or moisture content in the soil, which is crucial for rice cultivation, requiring flooded fields. Conversely, BSI reflects the soil condition. We incorporate these attributes into the model as their respective Mean and Variance features. Similar to the approach with vegetation indices, we have excluded the variance feature from the models due to its limited correlation with Yield. Notably, all water indices exhibit high correlations with each other and share a similar correlation with Yield.

Lastly, concerning meteorological characteristics, the average ambient air temperature (refers to variable ``$MET_{temp, mean}$") and specific humidity ($MET_{sh, mean}$) exhibit the strongest correlation with crop yield, and they also demonstrate a high degree of correlation between themselves. Solar radiation ($MET_{solrad, mean}$) emerges as a significant predictor due to its notable correlation with yield and comparatively lower correlation with other meteorological features.

\subsection{Feature Selection}
Up to this point, all the engineered features, totalling 94 in number, have undergone various stages of processing. These stages include (i) grouping, involving the arrangement of data types and condensation into categorical, numerical, and object types; (ii) scaling, which entails MinMax scaling; and (iii) splitting through a train-test split with a ratio of 3:1.

As outlined in the preceding sections, the subset of the 94 features exhibits significant correlation, and incorporating all these features in the models would lead to computationally intensive experiments. To mitigate this, during the final processing stage, we execute multiple rounds of feature selection, including Pairwise Feature Independence Check using the \(\chi^2\) test, statistical significance tests based on p-values, outlier removal, and thresholding for correlation and variance. Following these stages, the outcome is a refined set of 15 predictive (11 numerical \& 4 categorical) features and 1 target feature all of which are shown in Table \ref{tab:features}.

\begin{table}[ht!]
\centering
\caption{Engineered and extracted Features after selection stages.}
\begin{tabularx}{\textwidth}{CCp{2.5cm}p{2.5cm}}
\toprule
\textbf{Variable} & \textbf{Description} & \textbf{Type} & \textbf{Source} \\
\toprule
Season$_{\text{Enc}}$ & Crop season indicator & Categorical & Field \\
Dist$_{\text{Chau Phu}}$ & Crop location indicator & Categorical & Field \\
Dist$_{\text{Chau Thanh}}$ & Crop location indicator & Categorical & Field \\
Dist$_{\text{Thoai Son}}$ & Crop location indicator & Categorical & Field \\
$Yield_{kg}$ & Rice yield in kg at a specific point & Numerical & Field \\
$Rainfall_{growth, max}$ & Max rainfall growth in mm per hour & Numerical & NASA GES DISC \\
$Rainfall_{growth, sum}$ & Sum rainfall growth in mm per hour & Numerical & NASA GES DISC \\
$Rainfall_{maturity, max}$ & Max rainfall maturity in mm per hour & Numerical & NASA GES DISC \\
$VV_{mean}$ & Mean SAR image intensity in VV polarisation & Numerical & Sentinel-1 \\
$B08_{max}$ & Max NIR spectral band intensity & Numerical & Sentinel-2 \\
$RGVI_{mean}$ & Mean spectral index & Numerical & Sentinel-2 \\
$kNDVI_{mean}$ & Mean spectral index & Numerical & Sentinel-2 \\
$GCC_{mean}$ & Mean spectral index & Numerical & Sentinel-2 \\
$LST_{mean}$ & Mean land surface temperature & Numerical & Sentinel-3 \\
$MET_{solrad, mean}$ & Mean solar radiation & Numerical & Sentinel-3 \\
\hline
\textbf{$Yield_{rate}$} & \textbf{Rice yield in kg per hectare ($kg/ha$)} & \textbf{Target} & \textbf{Field} \\
\bottomrule
\end{tabularx}
\label{tab:features}
\end{table}

\begin{figure}[t!]
\centering
\includegraphics[width=\linewidth]{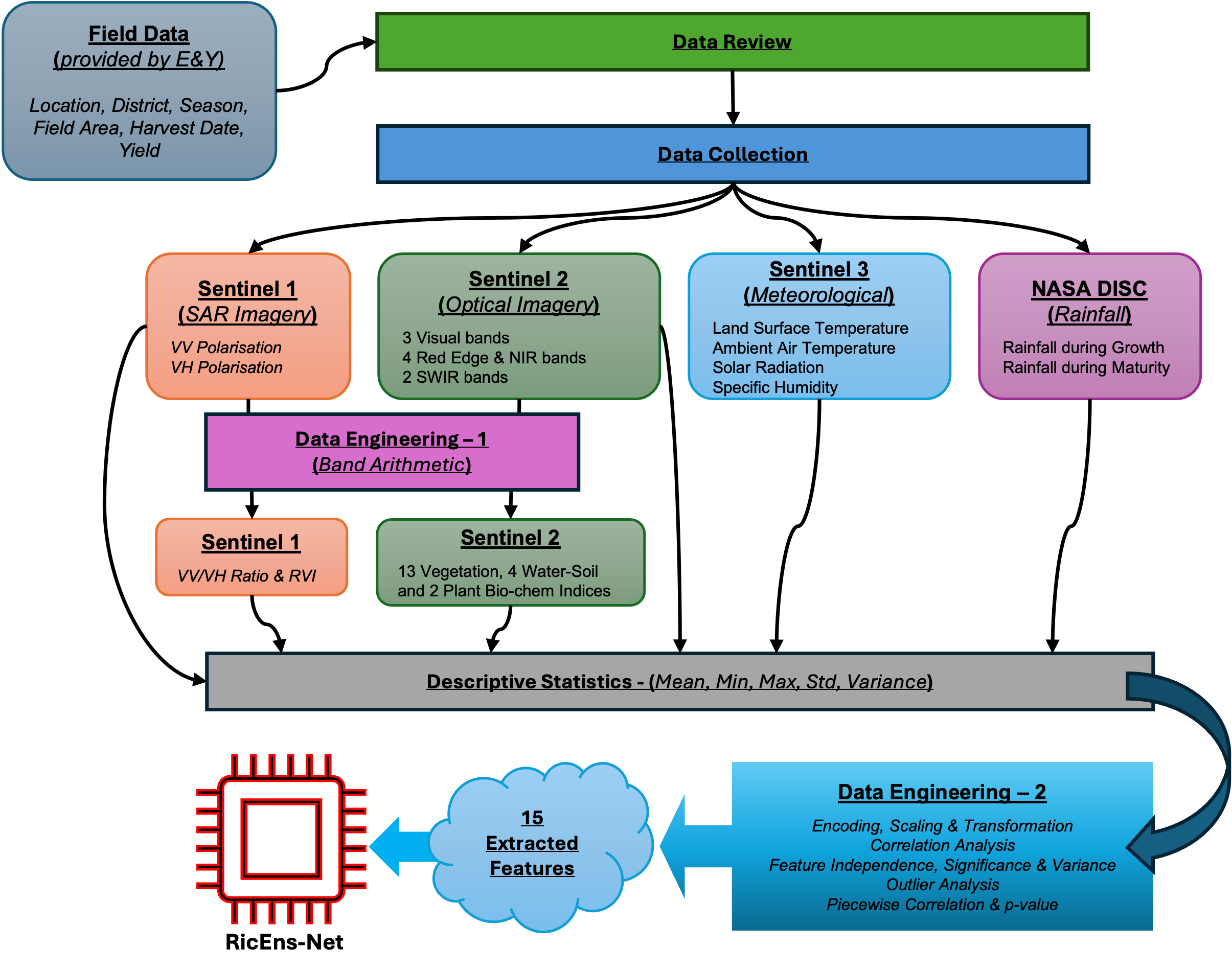}
\caption{Data collection, processing and engineering stages (Please see Figure \ref{fig:data2} and \ref{fig:ricensnet} for the details of the Data Engineering - 2 block and RicEns-Net model, respectively.).}
\label{fig:data}
\end{figure}

\begin{figure}[t!]
\centering
\includegraphics[width=\linewidth]{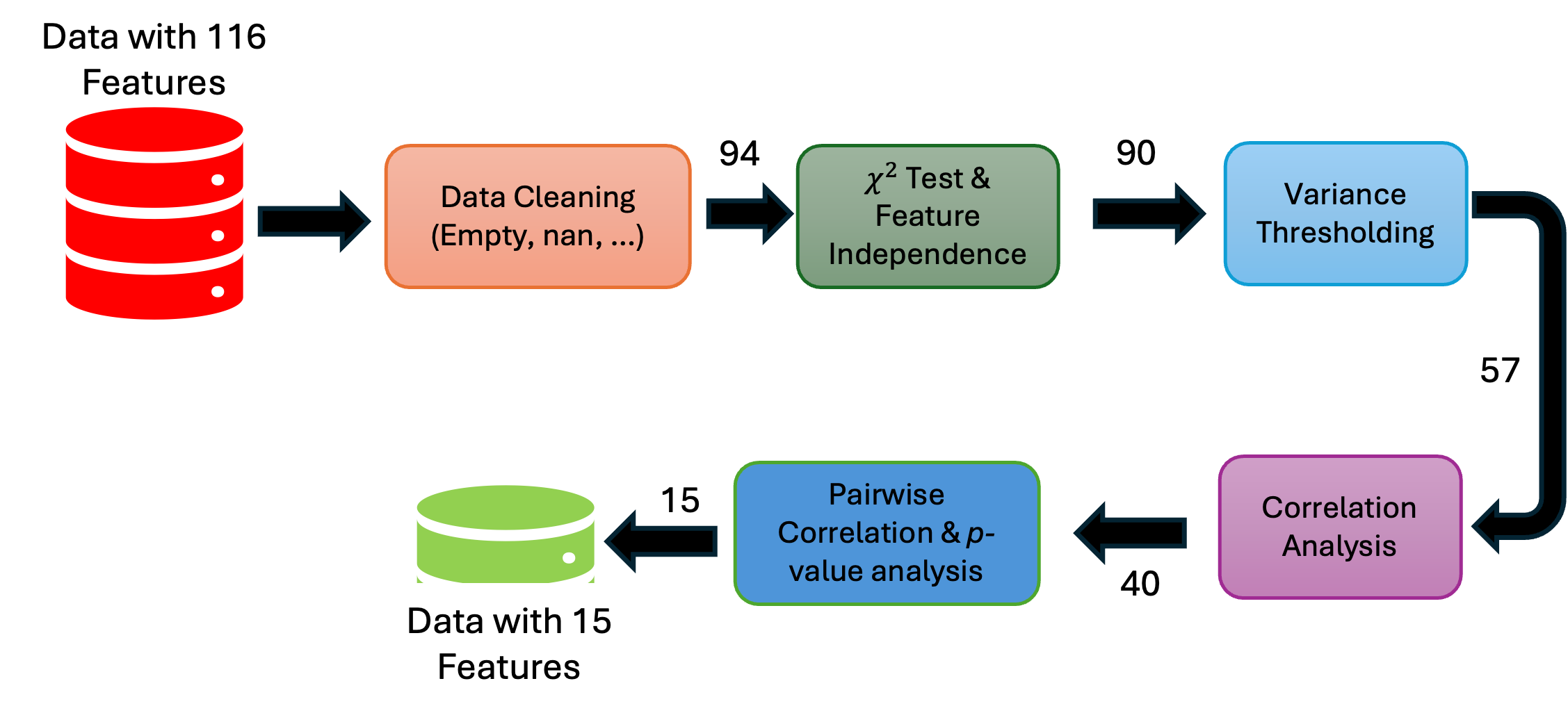}
\caption{A breakdown of the Data Engineering - 2 block.}
\label{fig:data2}
\end{figure}

The whole data collection, processing and engineering stages are plotted in Figure \ref{fig:data} whilst a detailed breakdown block diagram of the Data Engineering - 2 is also presented in Figure \ref{fig:data2}.

\subsection{The Proposed Model - RicEns-Net}
\begin{figure}[t!]
\centering
\includegraphics[width=\linewidth]{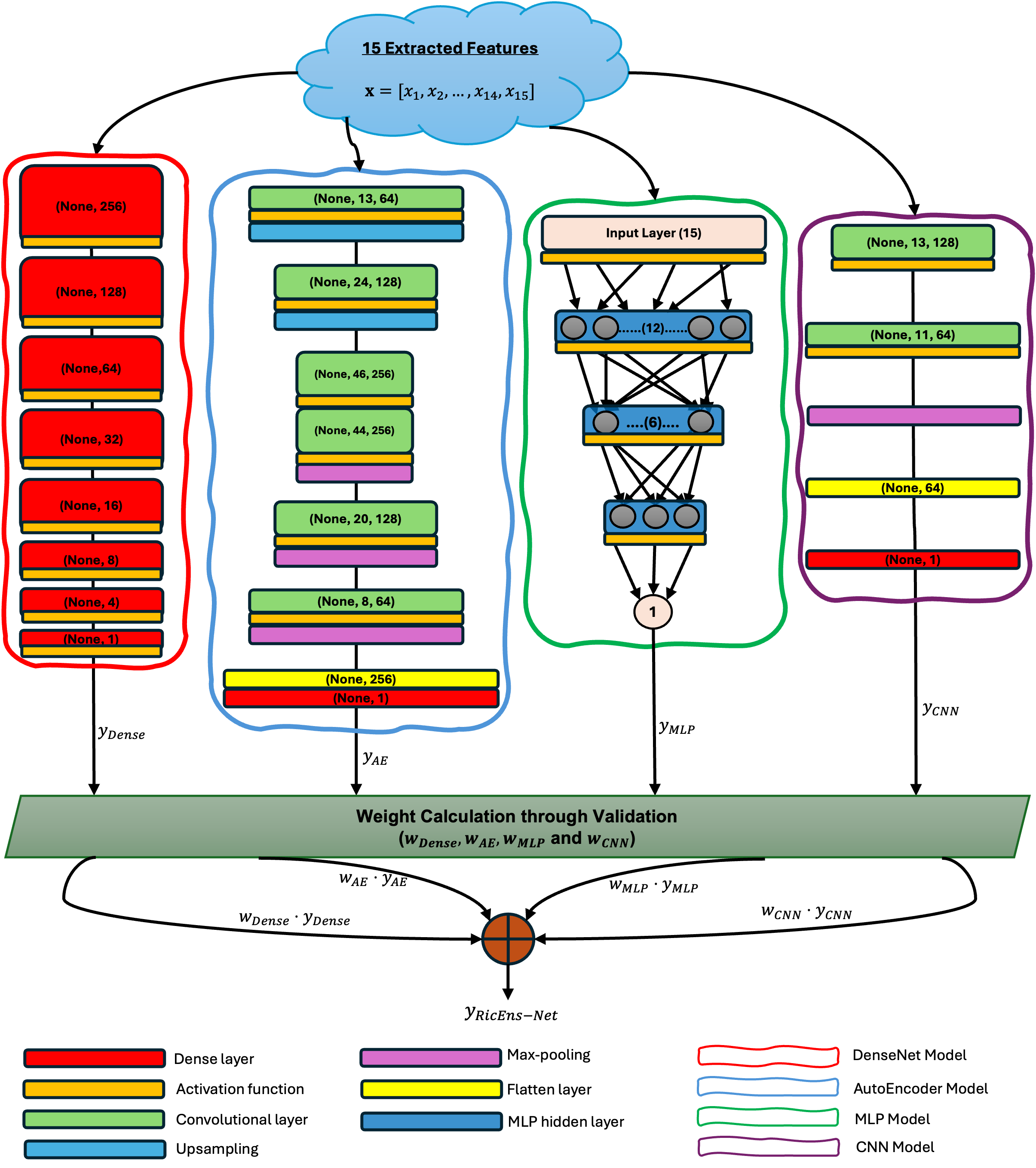}
\caption{RicEns-Net model details.}
\label{fig:ricensnet}
\end{figure}

After the extensive stages of data engineering, pre-processing, and preparation mentioned earlier, the 15 most informative and significant data features from 5 different data modalities will be employed to predict rice yield for the specified locations. This section provides a comprehensive introduction to the proposed deep ensemble model, \textit{RicEns-Net}. As mentioned in the earlier stages, deep learning models have currently been dominating yield prediction studies. This paper stands parallel with these advances in the literature but tries to explore complementary advantages of different deep learning regression techniques under a deep ensemble architecture named RicEns-Net. The details of RicEns-Net model architectures are presented in Figure \ref{fig:ricensnet}.

RicEns-Net utilises two traditional deep learning architectures, CNN and MLP. Despite the various designed versions of these models in the literature, no single architecture has emerged as a general solution. This motivated us to create our architecture tailored to achieve optimal performance for the rice yield prediction dataset. In addition to CNN and MLP, RicEns-Net incorporates two significant deep learning architectures: DenseNet and AutoEncoder (AE). While these architectures are scarcely used in yield prediction literature, they are prevalent in crucial remote sensing applications, such as semantic segmentation and classification.

Our motivation for incorporating DenseNet and AE architectures into our ensemble model stems from their respective advantages in handling complex data. AE architectures are valuable for rice yield prediction as they efficiently reduce dimensionality, extracting essential features while minimizing noise. Conversely, DenseNet architectures enhance deep learning by ensuring robust gradient flow through dense connections, thereby improving feature propagation and enabling the model to learn intricate patterns within agricultural data. These combined capabilities make the ensemble model more effective in predicting rice yield.

In developing the RicEns-Net ensemble model, extensive testing was conducted to determine the optimal architectures for each individual deep learning model. This rigorous process involved systematically evaluating various configurations, including the number of layers, types of activation functions, and loss functions. For each model, we experimented with different depths ranging from shallow to deep architectures to find the optimal balance between complexity and performance. We also tested various activation functions, such as ReLU, sigmoid, SiLU, and tanh, to identify the most effective function for each model. Additionally, we compared multiple loss functions, including mean squared error, mean absolute error, and Huber loss, to select the one that minimized prediction errors most effectively. Furthermore, each model's architecture was refined through a series of experiments, incorporating cross-validation and hyperparameter tuning, to ensure the best performance for predicting rice yield. This exhaustive testing process ensured that each model within the ensemble was optimally configured to contribute to the overall predictive power of RicEns-Net. The details of these architectures, including the final selected configurations, are illustrated in Figure \ref{fig:ricensnet}. These configurations were chosen based on comprehensive performance evaluations, ensuring that each component model enhances the proposed deep ensemble's accuracy and robustness.

In order to create the RicEns-Net's ensemble output, we use a weighted average approach. Let \( y_i \) be the output of the \( i \)-th model, and let \( w_i \) be the weight assigned to the \( i \)-th model. The ensemble output \( y_{RicEns-Net} \) is then given by:
\begin{align}
y_{\text{RicEns-Net}} = \sum_{i=1}^{N} w_i y_i
\end{align}
where \( N \) is 4 - the total number of models in the ensemble - and $i$ refers to each model with keywords $\{Dense, AE, MLP, CNN\}$.

The weights \( w_i \) are determined based on the validation errors of each model. Let \( e_i \) be the validation error of the \( i \)-th model. We assign the weights such that a lower validation error corresponds to a higher weight. Specifically, we can define the weight \( w_i \) as:
\begin{align}
w_i = \frac{\frac{1}{e_i}}{\sum_{j=1}^{N} \frac{1}{e_j}}
\end{align}
This choice ensures that the weights are normalised to sum to 1:
\begin{align}
\sum_{i=1}^{N} w_i = 1
\end{align}
Thus, the final ensemble output is a weighted average of the individual model outputs, where the weights are inversely proportional to the validation errors, normalised to sum to 1. This approach prioritises models with lower validation errors, giving them a higher influence on the ensemble output.

\subsection{On Deciding State-of-the-Art}
Despite significant advancements in computer vision research through the development of novel deep learning architectures, yield prediction research predominantly relies on traditional deep learning methods like CNN \cite{gavahi2021deepyield, boppudi2024deep, hashemi2024yield} and MLP \cite{maimaitijiang2020soybean, wu2024wheat, hussain2024analysing}, as well as state-of-the-art machine learning algorithms such as RF \cite{shahhosseini2021corn,abdali2024parallel,manafifard2024new}, XGBoost \cite{shahhosseini2020forecasting, shahhosseini2021corn, li2024global}, and SVR/SVM \cite{chen2022predicting,tao2023yield,abdali2024parallel}. According to a recent (2024) systematic review in \cite{trentin2024tree}, nearly one-third of the papers published in this area propose one of RF, SVM/SVR or CNN as the best-performing models. Additionally, traditional time series deep learning techniques, including Long Short-Term Memory (LSTM) \cite{boppudi2024deep, gopi2024red} and its variants, have been extensively utilized in the literature. Given this context, selecting the appropriate state-of-the-art models for the comparison study in this paper is of paramount importance. This section is dedicated to explaining the rationale behind the choice of each comparison model.

As previously discussed, this paper utilizes a dataset from the EY Open Science Challenge 2023. This global data science competition, sponsored by EY, a leading consulting firm, is notable for its innovative application of Data Science and Analytics to real-world business scenarios. The EY Open Science Challenge 2023 ran for two months, from January 31, 2023, to March 31, 2023, attracting over 13,000 participants and more than 7,500 submissions. The competition awarded USD 10,000 to the winner and USD 5,000 to the runner-up. EY has provided performance results for the winning teams (global and employees) on the same test dataset, making these two models ideal candidates for comparison in this study. The global winner used a CatBoost regression model (referred to as CatBoost-EY in this paper), while the employee winner's model was based on Extremely Randomized Trees (referred to as ExtRa-EY). We present these models' performance metrics without infringing on their copyrights.

Considering that our ensemble model incorporates four deep learning architectures— DenseNet, CNN, MLP, and AE— we also evaluate their individual performance in the comparison study. This approach aligns with the systematic review paper \cite{joshi2023remote}, which notes that 78 out of 102 proposed models in the literature up to 2023 include these four architectures. This validation supports our choice of comparison models, ensuring that the proposed RicEns-Net is benchmarked against state-of-the-art standards. Additionally, we incorporate advanced machine learning algorithms such as XGBoost, RF, SVR, AdaBoost, ElasticNet, and Gradient Boosting into the comparison pool, along with their Voting and Stacked ensemble models \cite{keerthana2021ensemble, abdali2024parallel}.

\subsection{Implementation of Models}
We conducted our research using Python version 3.10.9, leveraging the rich ecosystem of libraries available for data science and machine learning. Our desktop workstation, featuring a robust 20-core processor and ample 32GB of RAM, provided the computational power necessary for handling large datasets and complex modelling tasks efficiently. The widely-used \emph{scikit-learn} (\emph{sklearn}) module served as the cornerstone for implementing traditional and state-of-the-art machine learning algorithms, offering a comprehensive suite of tools for data preprocessing, model selection, and evaluation. Additionally, for the implementation of deep learning architectures, we employed the widespread Python libraries \emph{tensorFlow} and \emph{keras}, which provide powerful abstractions and efficient computation frameworks tailored specifically for neural network development. These Python modules enabled us to explore a diverse set of modelling techniques and methodologies, ultimately facilitating the realisation of our research objectives with precision and scalability.

The dataset was divided into training and test sets with a split ratio of 3:1. The training set comprised 75\% of the data, while the remaining 25\% was reserved for testing. To ensure robustness in the model evaluation process, a 10-fold cross-validation (CV) procedure was applied exclusively to the training data. During this stage, the training data was split into 10 subsets, with each subset used as a validation set once while the model was trained on the remaining 9 subsets. This process was repeated 10 times to account for variability in training, and the results were averaged to minimise bias and variance. The test data, which was kept entirely separate, was evaluated only once after the training phase to assess the final model performance. This approach ensures that the test set remains unbiased by the training process and provides a reliable estimate of the model's generalisation capability.

\subsection{Evaluation Metrics}
All the models will undergo assessment utilising regression metrics such as RMSE (Root Mean Square Error), MAE (Mean Absolute Error), $R^2$ Score (Coefficient Of Determination), and Adjusted $R^2$ (Adjusted Coefficient Of Determination). While MAE stands as the most straightforward performance metric, RMSE poses a more rigorous criterion by squaring the prediction error before calculating its mean and taking the square root. MAE and RMSE exhibit differences in their sensitivity to outliers.
\begin{align}
MAE & = \dfrac{1}{N}\sum\left(y_{predicted} - y_{actual}\right) \\
RMSE & = \sqrt{\dfrac{1}{N}\sum\left(y_{predicted} - y_{actual}\right)^2}
\end{align}

Another critical metric is $R^2$ which is anticipated to fall within the 0 to 1 range, although it can dip below 0 for specific models. In simpler terms, the $R^2$ value gauges the model's capacity to elucidate the variance of the target variable. To provide a more precise definition, as outlined in the SkLearn user guide, it signifies the proportion of variance (of $y$) explained by the independent variables in the model. It offers insight into the goodness of fit and serves as an indicator of how effectively the model is likely to predict unseen samples through the explained variance proportion \citep{scikit-learn-r2-score}.
\begin{align}
R^2 = 1 - \dfrac{\sum\left(y_{predicted} - y_{mean}\right)^2}{\sum\left(y_{actual} - y_{mean}\right)^2}
\end{align}

Additionally, we employ a modified version of $R^2$ known as Adjusted $R^2$, which factors in the impact of an elevated number of predictors contributing to a higher $R^2$ value. Furthermore, we compute the $R^2$ score for the input training data to compare it with the testing $R^2$ score. The disparity between them signals the potential for overfitting. The evaluation of models based on disparities in training and testing $R^2$ scores aids in the identification of models that may not generalise effectively with unseen data.
\begin{align}
Adjusted\text{ }R^2 = 1 - \left[ \dfrac{(1-R^2)\cdot(n-1)}{(n-k-1)} \right]
\end{align}
where $n$ refers to number of observations and $k$ is the number of predictors.

\begin{table}[htbp]
  \centering
  \caption{Rice crop yield prediction performance for all the utilised models. Bold and underlined values belong to the best and second-best models, respectively.}\label{tab:results}
    \begin{tabularx}{\textwidth}{Cp{1.25cm}p{1.25cm}p{1.25cm}Cp{1.25cm}p{1.25cm}p{1.5cm}}
    \toprule
    \textbf{Models} & \textbf{MAE} & \textbf{RMSE} & \textbf{Train $R^2$} & \textcolor{blue}{\textbf{10-fold CV Avg.} $R^2$} &\textbf{Test $R^2$} & \textbf{$R^2$ Diff.} & \textbf{Test Adj. $R^2$} \\
    \toprule
    XGBoost & 365.455 & 458.963 & 0.702 & \textcolor{blue}{0.666 $\pm$ 0.055} & 0.594 & 0.108 & 0.545 \\
    RF    & 360.824 & 449.734 & 0.733 & \textcolor{blue}{0.678 $\pm$ 0.059} & 0.610 & 0.123 & 0.563 \\
    GB    & 354.215 & 452.787 & 0.814 & \textcolor{blue}{0.656 $\pm$ 0.057} & 0.605 & 0.209 & 0.557 \\
    SVR   & \underline{345.711} & 448.902 & 0.689 & \textcolor{blue}{0.670 $\pm$ 0.051} & 0.612 & 0.077 & 0.565 \\
    ElasticNet & 351.660 & 450.985 & 0.688 & \textcolor{blue}{0.668 $\pm$ 0.058} & 0.608 & 0.080 & 0.561 \\
    CNN   & 349.048 & \underline{442.300} & 0.650 & \textcolor{blue}{\underline{0.688 $\pm$ 0.075}} & 0.623 & \textbf{0.027} & \underline{0.578} \\
    AdaBoost & 357.625 & 445.593 & 0.742 & \textcolor{blue}{0.679 $\pm$ 0.049} & 0.618 & 0.125 & 0.571 \\
    MLP   & 365.360 & 458.476 & 0.702 & \textcolor{blue}{0.665 $\pm$ 0.055} & 0.595 & 0.107 & 0.546 \\
    Voting & 351.604 & 446.990 & 0.779 & \textcolor{blue}{0.633 $\pm$ 0.026} & 0.615 & 0.164 & 0.569 \\
    Stacking & 375.251 & 467.695 & 0.691 & \textcolor{blue}{0.668 $\pm$ 0.064} & 0.579 & 0.112 & 0.528 \\
    DenseNet & 348.831 & 455.805 & 0.699 & \textcolor{blue}{0.642 $\pm$ 0.075} & 0.600 & 0.099 & 0.551 \\
    AE    & 360.050 & 453.462 & 0.680 & \textcolor{blue}{0.658 $\pm$ 0.100} & 0.604 & 0.076 & 0.556 \\
    \midrule
    CatBoost-EY & - & \underline{441.200} & - & - &- &- &-\\
    ExtRa-EY & 367.000 & 449.900 & - &- & - &- &-\\
    \midrule
    \midrule
    \textbf{RicEns-Net} & \textbf{341.125} & \textbf{436.258} & 0.696 & \textcolor{blue}{\textbf{0.692 $\pm$ 0.045}} & 0.633 & \underline{0.063} & \textbf{0.589} \\
    \bottomrule
    \end{tabularx}%
\end{table}

\begin{figure}[ht]
    \centering
    \includegraphics[width=\linewidth]{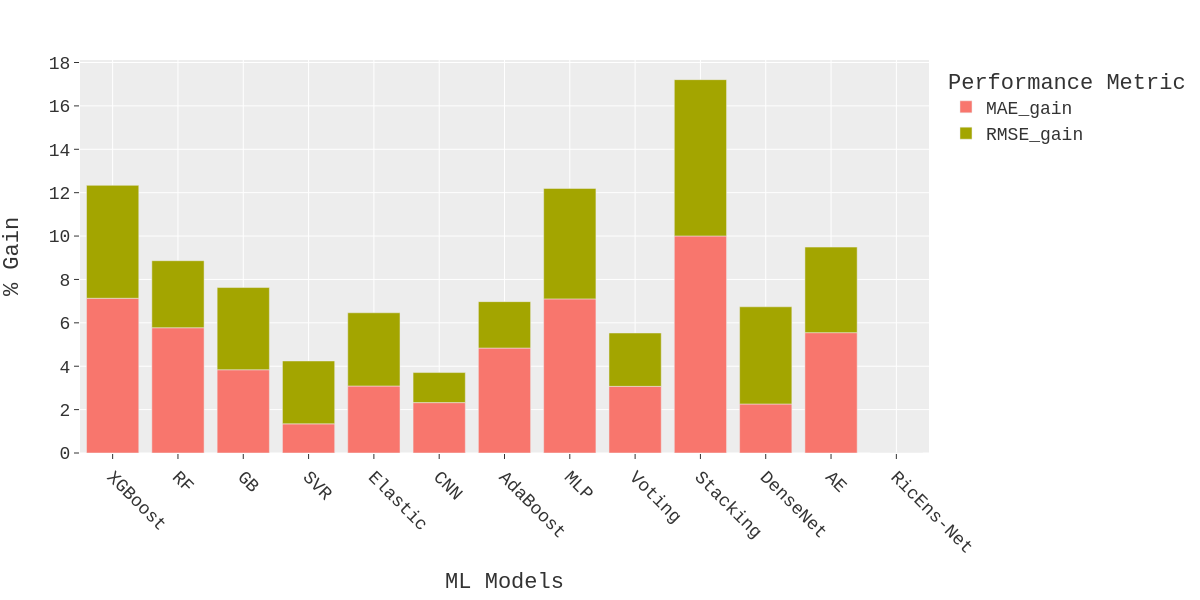}
    \caption{Percentage performance gain of the RicEns-Net model compared to all other reference models in terms of RMSE and MAE.}\label{fig:res1}
\end{figure}

\begin{figure}[ht]
    \centering
    \includegraphics[width=\linewidth]{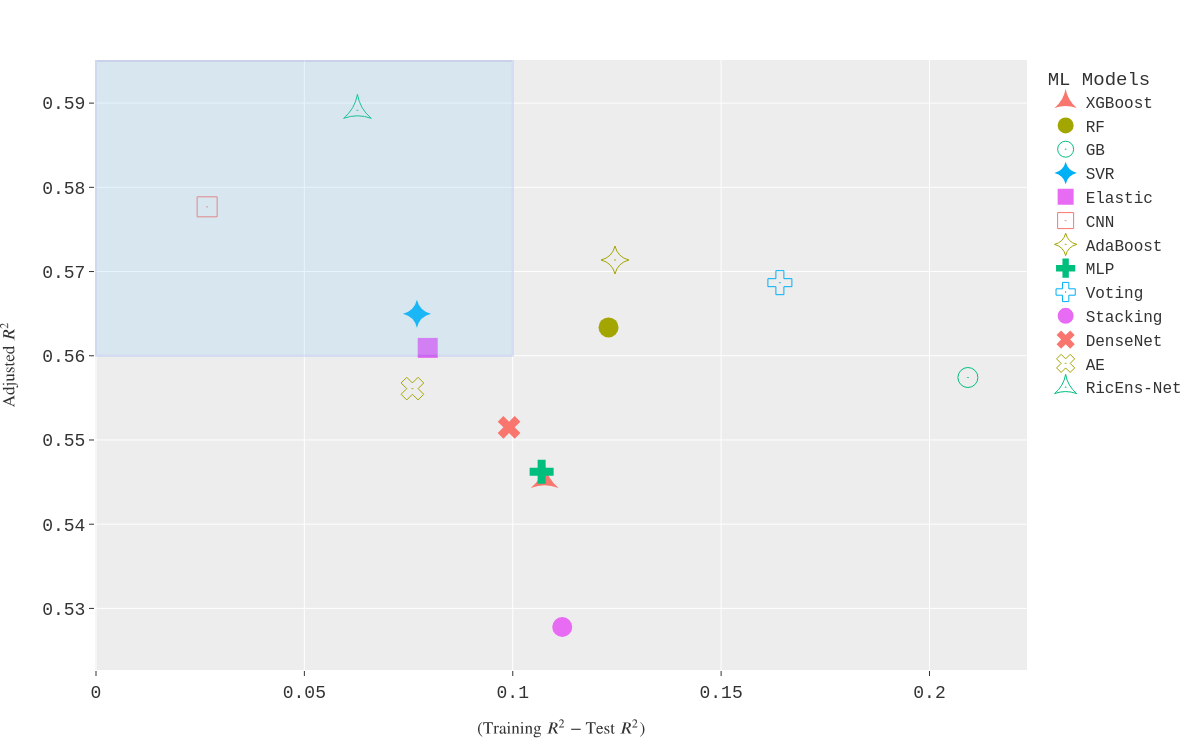}
    \caption{2D scatterplot of the Adjusted Test R2 and R2 difference values}\label{fig:res2}
\end{figure}

\begin{figure}[htbp]
    \centering
    \includegraphics[width=0.495\linewidth]{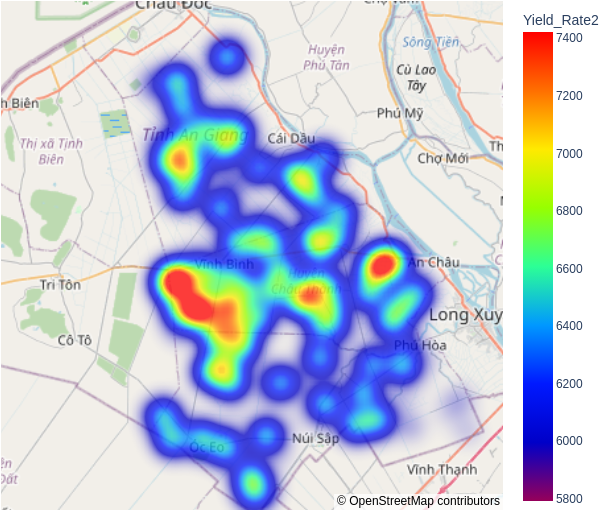}
    \includegraphics[width=0.495\linewidth]{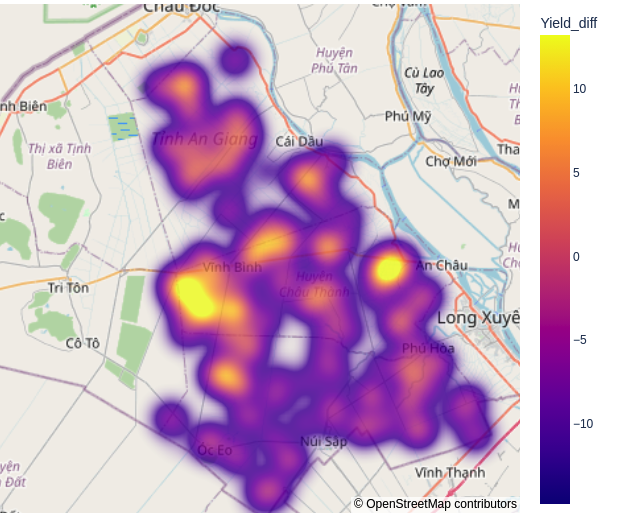}
    \caption{Map representation of the Deep Ensemble $Yield_{rate}$ (LEFT) predictions and, (RIGHT) prediction errors for the test set.}\label{fig:res34}
\end{figure}

\begingroup
\section{Results}\label{sec:results}
We begin by comparing the performance of all state-of-the-art models mentioned above against the proposed RicEns-Net model. Table \ref{tab:results} presents the MAE and RMSE errors in predicting rice yield, along with the goodness of fit measures including train/test and adjusted $R^2$ values.

Analysing the performance metrics in Table \ref{tab:results}, it is evident that the models evaluated for rice crop yield prediction exhibit a range of performance across multiple metrics, including MAE, RMSE, and adjusted $R^2$. Among these models, the proposed Deep Ensemble model, RicEns-Net, consistently demonstrates more precise predictive capability across all key indicators. Specifically, the RicEns-Net model achieves an MAE of 341.125 Kg/Ha and an RMSE of 436.258 Kg/Ha, representing improvements of around 2-5\% over the next best-performing models (CNN and SVR), depending on the metric.

RicEns-Net's standout performance in terms of adjusted $R^2$, where it achieves a value of 0.589, indicates a better overall fit compared to other models. This is particularly important as adjusted $R^2$ accounts for the number of predictors in the model, making it a critical measure of the model's generalization ability. By contrast, while traditional models such as Voting Regressor and CNN show relatively strong results, with adjusted $R^2$ values of 0.569 and 0.578 respectively, they still fall short of RicEns-Net's performance.

Furthermore, SVR and DenseNet, which also rank high in predictive performance, show RMSE values of 448.902 and 455.805 Kg/Ha, and MAE values of 345.711 and 348.831 Kg/Ha, respectively. However, despite their solid performance, they trail behind RicEns-Net by a margin of approximately 3\% in MAE and 4-5\% in RMSE.

In the evaluation of predictive performance across models, we performed a 10-fold cross-validation and calculated the average $R^2$ values for each approach. RicEns-Net demonstrated the highest $R^2$ score of 0.692 $\pm$ 0.045, outperforming other models in terms of predictive accuracy. Among the traditional machine learning models, RF and AdaBoost exhibited strong performance, with $R^2$ values of 0.678 $\pm$ 0.059 and 0.679 $\pm$ 0.049, respectively. Similarly, CNN achieved a competitive $R^2$ score of 0.688 $\pm$ 0.075, further validating the efficacy of deep learning architectures in this task. In contrast, models such as Voting Regressor and AE yielded lower $R^2$ scores of 0.633 $\pm$ 0.026 and 0.642 $\pm$ 0.075, indicating limitations in their ability to handle the complexity of multi-modal data. 

Another key observation is the variance in $R^2$ differences between the train and test sets for each model, highlighting the generalization strength of each approach. RicEns-Net has one of the lowest $R^2$ differences at 0.063, indicating minimal overfitting compared to other models such as Gradient Boosting (0.209) and Voting (0.164). This is a significant achievement, as it suggests that RicEns-Net maintains stable performance across both training and unseen test data, further reinforcing its reliability for real-world applications.

While models like XGBoost, RF, and Gradient Boosting also exhibit reasonable performance, with RMSE values ranging from 449.734 to 458.963 Kg/Ha, and MAE values between 354.215 and 365.455 Kg/Ha, they are consistently outperformed by RicEns-Net in all metrics. Notably, CatBoost-EY, one of the models from the EY Data Science Challenge, shows a strong RMSE value of 441.200 Kg/Ha but lacks data for other metrics, making a full comparison difficult. On the other hand, when comparing the performance of RicEns-Net with the EY Open Science Challenge winning models, the results in Table \ref{tab:results} show that RicEns-Net consistently outperforms these models in terms of MAE and RMSE. Furthermore, models such as CNN, SVR, DenseNet, and Voting Regressor also display competitive performance compared to the EY winners.

In addition to the quantitative analysis of model performances, Figures \ref{fig:res1}, \ref{fig:res2}, and \ref{fig:res34} provide visual representations that further elucidate the effectiveness of RicEns-Net compared to other models.

\textbf{Figure \ref{fig:res1}} presents a stacked bar chart comparing the MAE and RMSE values across all models. This visualization highlights the performance improvement of RicEns-Net over the alternatives. Notably, CNN, SVR, and the Voting Regressor models show relatively small performance degradation, with only around a 5\% drop in accuracy compared to RicEns-Net. These models are the closest competitors to RicEns-Net in terms of cumulative error metrics. However, models like the Stacking Regressor, MLP, and XGBoost exhibit more than a 10\% increase in errors, indicating significantly lower predictive accuracy. This graphical representation clearly underscores RicEns-Net's capability in minimizing both MAE and RMSE.

\textbf{Figure \ref{fig:res2}} provides a scatter plot that illustrates the relationship between the $R^2$ Difference (train-test gap) and the adjusted test $R^2$ values, offering insight into model generalization and fit. The most effective models, occupying the optimal top-left region of the scatter plot, include RicEns-Net, CNN, SVR, and ElasticNet. These models not only deliver high adjusted $R^2$ scores but also maintain small differences between their training and test performances, indicating minimal overfitting and robust generalization. In contrast, models like the Voting Regressor and AdaBoost, despite achieving top-five rankings in terms of MAE and RMSE, show a significant disparity between training and test $R^2$ values. This suggests that these models overfit the training data and fail to generalize as well to unseen data. RicEns-Net's placement in the optimal region confirms its strong generalization ability and balanced performance across training and test sets.

Finally, \textbf{Figure \ref{fig:res34}} offers a detailed visual analysis of RicEns-Net's predictions for crop yield ($Yield_{rate}$), focusing on spatial error distribution across the test dataset. This figure reveals that, while RicEns-Net maintains consistently low error margins across most of the test regions, there are small areas where the maximum absolute error exceeds 10\%. These outlier regions are concentrated in areas with higher actual yield rates, where RicEns-Net tends to underestimate $Yield_{rate}$. This observation suggests that while RicEns-Net excels in overall yield prediction, further refinement could be beneficial for regions with higher yields. Nevertheless, the majority of the test areas exhibit low prediction errors, further validating the model's reliability in estimating crop yield under diverse conditions.

To sum up, the combined results from these figures highlight RicEns-Net's competitive advantage, not only in terms of quantitative metrics but also in its ability to generalize well across different data conditions and accurately predict crop yield with minimal errors across various regions.

\section{Discussion}\label{sec:disscussions}
The results demonstrate the effectiveness of the proposed RicEns-Net model in achieving high accuracy for crop yield prediction. With a 2\% to 10\% improvement in both MAE and RMSE compared to other state-of-the-art models, RicEns-Net clearly shows its capability, particularly in handling multi-modal data from radar, optical, and meteorological sources. This aligns with previous findings where deep ensemble models have demonstrated enhanced predictive capabilities through the combination of diverse feature sets \cite{abdali2024parallel, shahhosseini2020forecasting, keerthana2021ensemble}. The significant improvement in adjusted $R^2$ further emphasizes the robustness of RicEns-Net, which surpasses even advanced models such as CNN and SVR.

A closer inspection of the performance of traditional ensemble methods like the Voting Regressor reveals their strong performance in MAE and RMSE metrics. However, these models exhibit overfitting tendencies, as indicated by the large $R^2$ disparities between training and test phases. This overfitting issue highlights a major advantage of RicEns-Net, which maintains a balance between training and test performance, avoiding overfitting through its deep ensemble design.

The performance of the RicEns-Net model, achieving a mean absolute error (MAE) of 341 kg/Ha, demonstrates its robust predictive capability in the context of crop yield forecasting for the Mekong Delta region in Vietnam. This region, often referred to as Vietnam’s ``rice bowl," contributes more than half of the nation’s rice output and is critical to ensuring both domestic food security and Vietnam’s position as a leading global rice exporter \cite{FAO:2022,su12125123}. Given the region’s susceptibility to climate variability, such as temperature fluctuations and erratic rainfall patterns, achieving high predictive accuracy in yield estimates is essential for implementing timely interventions and optimizing resource allocation. The accuracy of 341 kg/Ha represents a meaningful advancement, given that the lowest average yield in the region is approximately 6,000 kg/Ha, as reported by \cite{clauss2018estimating}. This error equates to around 5–6\% of the lowest yield, reaching a maximum of 10\% in some cases, as illustrated in Figure \ref{fig:res34}. These findings underscore the value of RicEns-Net in enhancing crop yield predictions and its potential for broader applications in precision agriculture.

Additionally, a 10-fold cross-validation was conducted to ensure the robustness of the results across different training and testing sets. The average $R^2$ values for RicEns-Net were the highest (0.692 $\pm$ 0.045) among all tested models, further validating its predictive skills. Other models like CNN and RF performed well, but RicEns-Net consistently outperformed them, confirming the advantage of deep ensemble methods in handling multi-modal data and mitigating overfitting risks.

Furthermore, when comparing the results to the EY Open Science Challenge 2023 winners, the competitive performance of RicEns-Net is evident. This further validates the efficacy of the proposed deep ensemble approach for predictive tasks relying on multi-modal data. While models like CNN and SVR also achieve competitive results, the effectiveness of neural network-based architectures for large-scale agricultural datasets is underscored. However, the underestimation of yield rates in high-yield regions, as noted in Figure \ref{fig:res34}, suggests that future work could explore refining the model for these specific outlier cases.

Despite the promising results, it is important to acknowledge several limitations and assumptions within this study. First, biases in the data could stem from the quality and availability of remote sensing and meteorological inputs. For instance, the presence of cloud cover in multispectral imagery may introduce spatial and temporal gaps in the dataset, affecting model training and predictions in certain regions. Furthermore, while the RicEns-Net model demonstrated robust performance across various environmental conditions, its generalizability to regions with different climate patterns, crop types, or farming practices remains to be fully tested. The model assumes that the selected features adequately capture the variability in yield, yet this may not account for site-specific factors such as soil health, irrigation methods, or pest control practices. Future research should aim to address these assumptions by expanding the dataset across diverse geographical and agronomic contexts and incorporating additional ground truth data to validate the model's broader applicability.

The practical implications of RicEns-Net for precision farming are significant. By accurately predicting crop yields based on multi-modal data, farmers can make more informed decisions regarding resource allocation, such as optimizing irrigation schedules and adjusting fertilizer inputs based on predicted yield potential. The use of radar and optical remote sensing data in the model allows for real-time monitoring of soil moisture and crop health, enabling farmers to act proactively to prevent yield losses due to drought or pest infestations. Furthermore, RicEns-Net’s ability to incorporate meteorological data provides valuable insights into how weather conditions might impact yields, allowing for better risk management and crop planning. As remote sensing technologies become more accessible, integrating models like RicEns-Net into farm management systems can offer farmers a user-friendly interface to visualize yield predictions and make data-driven decisions, ultimately leading to increased efficiency and sustainability in agricultural practices.

Another promising area for further research lies in enhancing the temporal resolution of the data, particularly in cases where satellite data acquisitions are limited by weather conditions. Leveraging UAVs equipped with hyperspectral and LiDAR sensors, which offer flexible scheduling and improved data collection even under adverse weather, can be a pivotal step toward refining yield prediction. Integrating these technologies into the existing multi-modal framework will further extend the utility of remote sensing in precision agriculture. Additionally, exploring the use of emerging machine learning models, such as transformers and self-supervised learning techniques, could lead to further advancements in predictive performance.

To sum up, this study contributes to the growing body of research on precision agriculture, demonstrating that deep ensemble models like RicEns-Net can significantly outperform traditional machine learning models and state-of-the-art techniques. Nevertheless, future research could investigate incorporating additional data sources or refining the feature engineering process to further improve the model’s performance, particularly in regions with extreme yield rates.

\section{Conclusions}\label{sec:conc}
In this study, we developed RicEns-Net, a novel deep ensemble model, for accurate rice yield prediction based on multi-modal data from remote sensing and meteorological sources. This approach proved highly effective, as demonstrated by the model’s greater performance in terms of key metrics such as RMSE and MAE. By employing advanced feature engineering and selection techniques, we identified 15 crucial features from radar, optical, and meteorological data, which allowed us to mitigate the curse of dimensionality and enhance prediction accuracy. The findings of this study suggest that RicEns-Net offers significant potential for real-world applications in crop yield forecasting, highlighting the utility of multi-modal data fusion in addressing agricultural challenges.

While RicEns-Net performed well, the limitations of cloud cover in optical remote sensing and the absence of field boundary delineation present opportunities for future improvements. Cloud detection algorithms, particularly those using Sentinel-2’s Band 9 for water vapour detection, should be integrated into the model to enhance the quality of spectral data in cloud-prone regions. Additionally, future research could focus on incorporating algorithms capable of mapping individual crop field boundaries to improve the granularity of yield predictions.

Looking ahead, emerging technologies such as UAVs equipped with multi-spectral, LiDAR, and radar sensors hold significant promise for enhancing the scope and accuracy of yield prediction models. These sensors can operate under variable weather conditions and at lower altitudes, offering flexibility in data collection. Furthermore, integrating biochemical and physiological crop monitoring techniques into multi-modal data fusion frameworks will push the boundaries of precision farming, offering novel insights into crop health and resource management. These future directions not only build upon the advancements demonstrated by RicEns-Net but also pave the way for the next generation of precision agriculture technologies.
\endgroup

\appendix
\section[\appendixname~\thesection]{Vegetation, Soil and Plant Biochemical Indices}
This appendix section presents important vegetation, soil and plant biochemical indices in Table \ref{tab:indices}.

\begin{table}[htbp]
\renewcommand{\arraystretch}{0.75}
    \centering
        \caption{Vegetation, Soil, Water and Plant Biochemical Indices}
    \label{tab:indices}\small
    \begin{tabular}{p{9.5cm}p{6.75cm}}
    \toprule
        \textbf{Index}	& \textbf{Reference}	\\\toprule
Normalized Difference Vegetation Index (\textbf{NDVI}) &	 $\dfrac{R_{NIR} - R_{RED}}{R_{NIR} + R_{RED}}$ \\\midrule
Transformed Vegetation Index (\textbf{TVI}) & $\sqrt{NDVI + 0.5}$\\\midrule 
Simple Ratio (\textbf{SR}) & $\dfrac{R_{NIR}}{R_{RED}}$\\\midrule
Enhanced Vegetation Index (\textbf{EVI}) & $2.5 \times \dfrac{R_{NIR} - R_{RED}}{R_{NIR} + 6R_{RED} - 7.5R_{}+1}$\\\midrule
EVI - 2-Bands (\textbf{EVI2}) & $2.5 \times \dfrac{R_{NIR} - R_{RED}}{R_{NIR} + 2.4R_{RED}+1}$\\\midrule
Soil adjusted vegetation index (\textbf{SAVI}) &	$1.5 \times \dfrac{R_{NIR} - R_{RED}}{R_{NIR} + R_{RED} + 0.5}$\\\midrule
Rice Growth Vegetation Index (\textbf{RGVI}) & $1 - \left(\dfrac{R_{} + R_{RED}}{R_{NIR}+R_{SWIR1}+R_{SWIR2}}\right)$\\\midrule
Difference Vegetation Index (\textbf{DVI}) & $R_{NIR} - R_{RED}$\\\midrule
Modified Simple Ratio (\textbf{MSR}) & $\dfrac{SR - 1}{\sqrt{SR+1}}$\\\midrule
Near Infra-Red Reflectance Of Vegetation (\textbf{NIRv}) & $NDVI \times R_{NIR}$ \\\midrule
Kernelized NDVI (\textbf{kNDVI}) & $\tanh{(NDVI^2)}$\\\midrule
NDVI-Red Edge (N\textbf{NDVIre}) & $\dfrac{R_{NIR} - R_{Red Edge1}}{R_{NIR} + R_{Red Edge1}}$\\\midrule
Normalized Difference Red Edge 1 (\textbf{NDRE1}) & $\dfrac{R_{Red Edge2} - R_{Red Edge1}}{R_{Red Edge2} + R_{Red Edge1}}$\\\midrule
Normalized Difference Red Edge 2 (\textbf{NDRE2}) & $\dfrac{R_{Red Edge3} - R_{Red Edge1}}{R_{Red Edge3} + R_{Red Edge1}}$\\\midrule
Normalized Difference Water Index (\textbf{NDWI}) 	&	 $\dfrac{R_{GREEN} - R_{NIR}}{R_{GREEN} + R_{NIR}}$\\\midrule 
Bare Soil Index (\textbf{BSI}) & $\dfrac{(R_{RED} + R_{SWIR1}) - (R_{NIR} + R_{})}{(R_{RED} + R_{SWIR1}) + (R_{NIR} + R_{})}$\\\midrule
Land Surface Water Index (1.6 $\mu$m) (\textbf{LSWI16}) & $\dfrac{R_{NIR} - R_{SWIR1}}{R_{NIR} + R_{SWIR1}}$\\\midrule
Land Surface Water Index (2.2 $\mu$m) (\textbf{LSWI22})& $\dfrac{R_{NIR} - R_{SWIR2}}{R_{NIR} + R_{SWIR2}}$\\\midrule
Chlorophyll Carotenoid Index (\textbf{CCI}) & $\dfrac{R_{GREEN} - R_{RED}}{R_{GREEN} + R_{RED}}$\\\midrule
Green Chromatic Coordinate (\textbf{GCC}) & $\dfrac{R_{GREEN}}{R_{RED} + R_{GREEN} + R_{}}$\\
\bottomrule
    \end{tabular}
\end{table}

\bibliographystyle{elsarticle-num}
\bibliography{elsarticle-template-num}
\end{document}